# Indonesia embraces the Data Science

Hokky Situngkir
(hokky.situngkir@surya.ac.id)
Bandung Fe Institute – Research Center for Complexity Sciences
Surya University

**Abstract**
The information era is the time when information is not only largely generated, but also vastly processed in order to extract and generated more information. The complex nature of modern living is represented by the various kind of data. Data can be in the forms of signals, images, texts, or manifolds resembling the horizon of observation. The task of the emerging data sciences are to extract information from the data, for people gain new insights of the complex world. The insights may came from the new way of the data representation (function of *R(x)* over data *x*), be it a visualizations, mapping, or other. The insights may also come from the implementation of mathematical analysis and/or computational processing (function *A(x)* over data *x*) giving new insights of what the states of the nature represented by the data. Both ways implement the methodologies reducing the dimensionality of the data. The relations between the two functions, *R(x)* and *A(x)* are the heart of how information in data is transformed mathematically and computationally into new information. The paper discusses some practices, along with various data coming from the social life in Indonesia becoming the variables within *R(x)* and *A(x)* to gain new insights about Indonesia in the emerging data sciences. The data sciences in Indonesia has made Indonesian Data Cartograms, Indonesian Celebrity Sentiment Mapping, Ethno-Clustering Maps, social media community detection, and a lot more to come, become possible. All of these are depicted as the exemplifications on how “data science” has become integral part of the technology bringing data closer to people.

## 1. Introduction

Today information technology has turned people's daily life into data generating processes. Life has never been so "recorded" and the net citizen has become data contributor as well as data consumer. While mathematics is the scientific language, the task of pointing out and speaking of patterns within data is demanding even more. When applied mathematics goes to information technology, the notion of "data science" is emerged [27]. Data science has become the emerging school of thought in the mix of applied mathematics, computation, and the substantive knowledge of the domain on which the data is resembled upon [2].

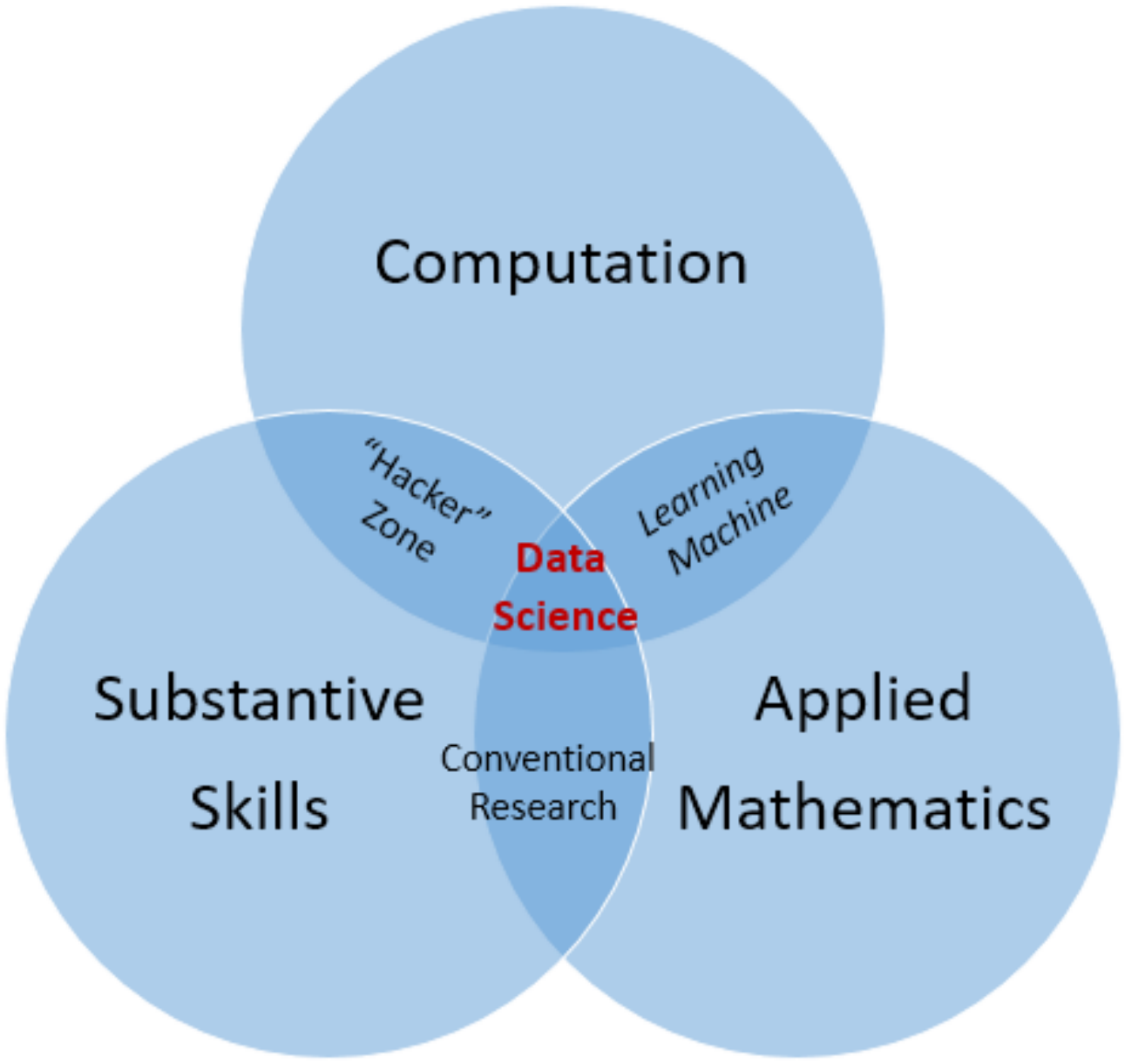

**Figure 1.** "Data Science" Venn diagram [*cf.* 2]

While the Data Science [1] emerges in the revolutionary information technology, the 270 million population living in the vast archipelagic geography of Indonesia has also become part of the data-generating society in need for Data Science. The horizon of data sciences of Indonesia is at large from the cultural heritage data to the political dynamics, from the unique archipelagic state to the sentiment among the diverse society, be it religions, moral politics, and economy. The role of mathematics is there with the possibility to formalize data and problems, with great deal of scientific methods and models, with which allows rigorous proofs of functional methodologies. The existing Data Science gives the ability to let Indonesian see Indonesia analytically, gives insights about themselves with their uniqueness, and solve the detected problems sharpened scientifically [*cf.* 14].

## 2. On method

Extracting information from the data is the mathematical process of data crunch. Say we have variables $x$, then the data crunch can be denoted as the process of making it into a sort of representation, $f: R(x)$, and or the process to use canon of analytical models reducing the complexity of the data [16], say $f: A(x)$. The task of science, whatsoever is to give simple view on the complex nature [11] of the approached objects. The nature of the data, $x$, is various. It depends on the substantive matter of the domain being approached. The processing of the data follows the form of the data. The data can be in the form of,

1. signals, thus $f: x \subset \mathbb{R} \rightarrow \mathbb{R}$.
2. images, thus $f: x \subset [0,1]^2 \rightarrow \mathbb{R}$
3. data in text, thus thus $f: x \subset \mathbb{L} \rightarrow \mathbb{R}$
4. data in manifolds, thus thus $f: x \subset S^2 \rightarrow \mathbb{R}$

From both ways, $R(x)$ and $A(x)$, observer may have insights from what the data is about. Mathematical models along with algorithms are used within both of the functions. However, in the whole process, the output variables of one function can be the input for the other. Some representations, $R(x)$, might have been simple enough for observer to look at into, but sometimes it needs to be processed again analytically for the clearer insights about the data, and *vice versa*. The general process can be drawn as in figure 2.

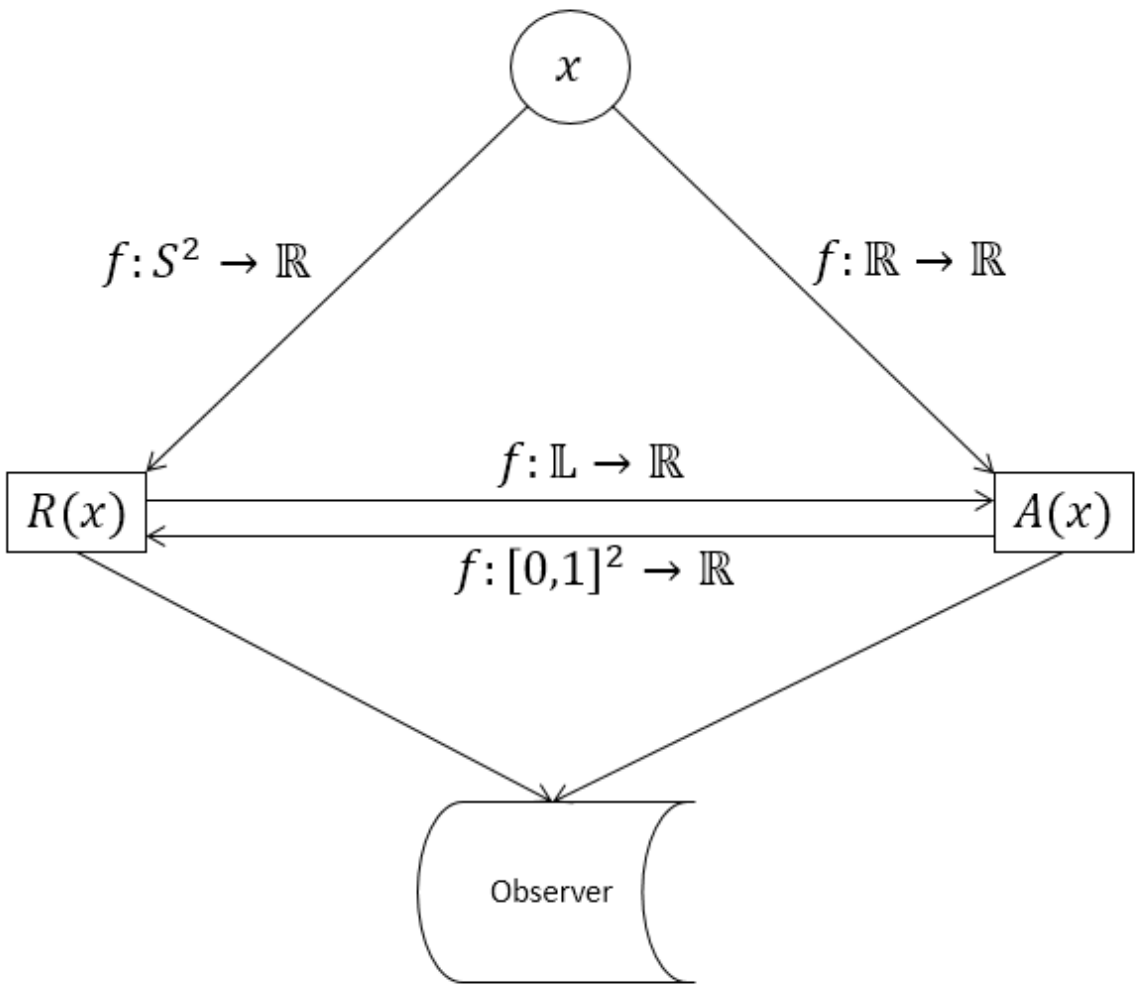

**Figure 2.** The interplay between data representation and analytics.

## 3. Review of Some Implementations

Data in the form of manifolds expressing geo-spatial information can be represented directly in the map or portrait of the surface of the earth. This kind of information is related to the geographical information system (GIS), and with some computational processes, is able to be delivered for observer demanding the spread of data on earth. Along with the cartography with geographical information system is an innovation related to the Geo-visualization statistics, name cartogram. Cartograms are a way to present data about some spatial areas on the soil land. The area on earth is depicted on proportionality with the statistics of the area. Thus, if we have the longitude ($x$) and latitude ($y$) points surrounding an area $G_{geographical_vectors}\{(x,y)\}$, calculations are delivered to transform the set of the geo-coordinates into vectors of cartograms,

$$R(S_{[x,y]}) \equiv G_{geographical_vectors}\{(x,y)\} \xrightarrow{T} K_{cartogram}\{(x,y)\} \quad (1)$$

where $T$ is the algorithm transforming the earth surface area due to the value of the statistics of the are,

$$\frac{\partial(T_x,T_x)}{\partial(x,y)} = \frac{\rho(x,y)}{<\rho>} \tag{2}$$

In the transformation, data density $\rho(x, y)$ is projected into the geographical area, by respect to the average of all density data $< \rho >$. In this fashion, the total area will be the same size before and after the transformation [3]. Figure 3 shows the example of political parties' exposures based on the result of Indonesian General Election 2014. The cartograms is drawn by using the population within regions and colored by the dominance of the political parties. Instantaneously, observer may get the idea on the popular votes among population in Indonesia [16].

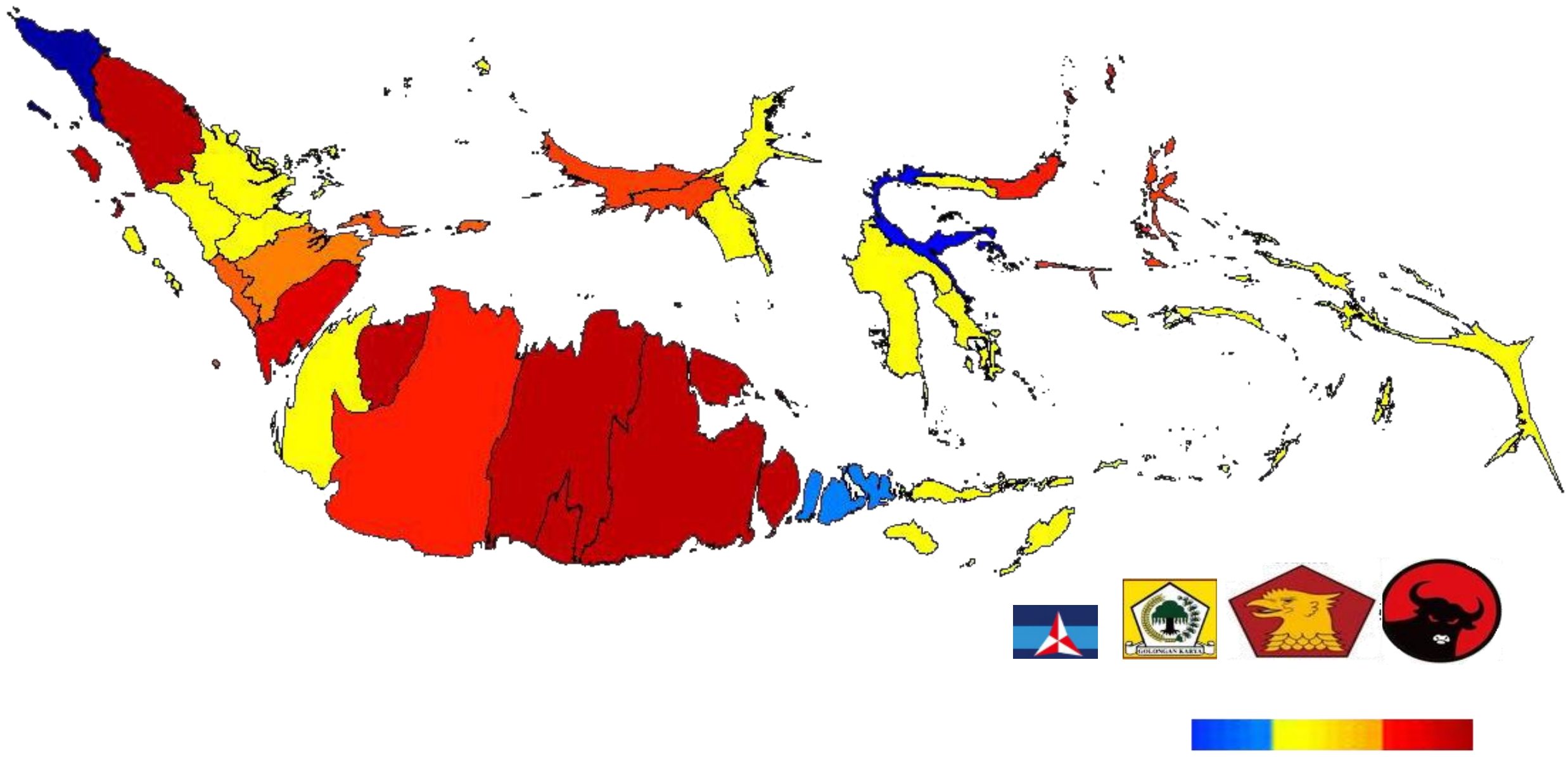

**Figure 3.** The cartograms of Indonesian population within the archipelago colored by the general election 2014 of the four biggest political parties [10].

The trend on using statistical mechanics model on economic and financial data is probably one of the most remarkable stories in data scientific community. This trend was introduced in the end of the previous century with the tagline "econophysics" [5, 7]. This trend uses models from statistical physics to see patterns in the high frequency data of finance [23] and later economic data. One thing that makes statistics different with data science, is about the finding of data distribution that is not normal on the dynamics of many prices in the market, which is power-law distribution [12]. While mostly statistics is using normal distribution as standard of approach [24, *cf.* 23], the practice of econophysics learn the data distribution from the data itself [8, *cf.* 26]. Many works have delivered insights about this, acquiring observation on seeing data in the form of signals [22].

In some cases, the analytical works on prices in the market as signals, can be delivered even further by transforming the collections of market data prices ($p \in \mathbb{R}$) into mapping. We can do this by transforming the correlation ($c_{xy}$) between prices of $x$ with the prices of $y$ into the distance in the ultrametric space [15],

$$R \circ A(p) \equiv d_{xy} = \sqrt{2(1 - c_{xy})} \tag{3}$$

The composition function above is exemplified with the prices of domestic needs as shown in figure 4. This visualization depicts the mapping of the ups and downs of the products' prices in Indonesian people's markets. The mapping somehow gives insight how changes of price of products correlated one another dynamically within daily basis in the market.

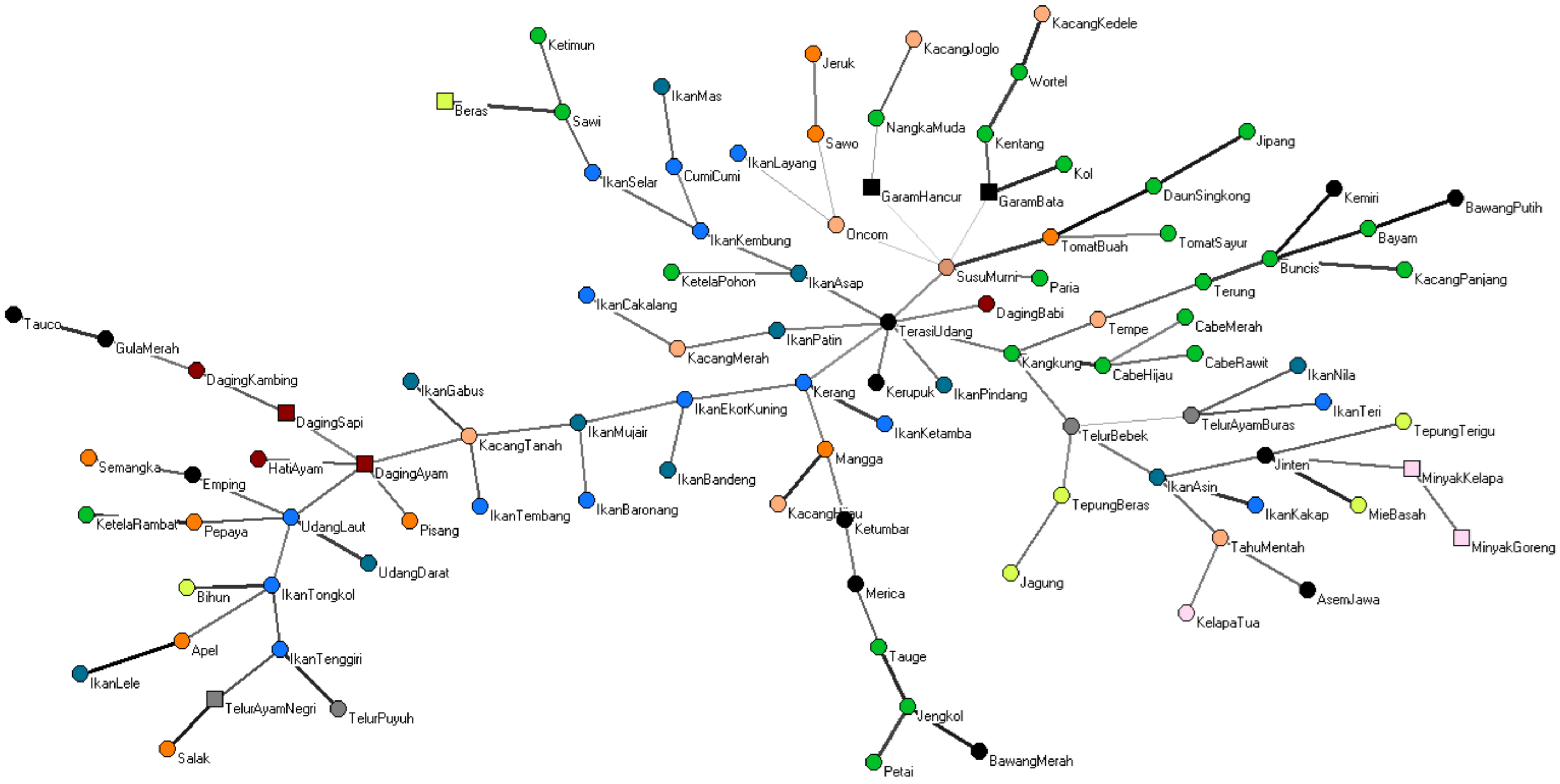

**Figure 4.** The correlative representation of domestic needs prices in Indonesian market 2000-2012 [26].

When it comes to the data series in the form of images, some image-related processing is in demand if we would like to serve the observer with mapping about large amount of data image. An interesting practice may be the mapping of Indonesian Batik.

Batik is a sort of traditional painting on fabric, and used mostly by Indonesian as traditional clothing. Almost all of ethnic groups in the archipelago have their own unique batik designs [18]. Research on thousands of Indonesian batik from all over the country by calculating the fractal dimension [17] and color histogram delivered some aspects of elementary geometry within batik. From the analysis, the 'quantitative differences' among batik is treated as homological vectors between one another [6]. The result is transformed into the matrix of Hamming distances, $\delta_{hamming}$, among those batik designs. This matrix is thus used to draw the mapping of Indonesian Batik, expressing the differences among batik aiming to show the interesting nature of the diversity emerged from the collective intelligence [13] in the archipelago. Due to our formalization, the process can be written as functional composition,

$$R \circ A([0,1]^2) \equiv m_{xy}(Img) \xrightarrow{T} \delta_{hamming} \tag{4}$$

The functions depicts the algorithmic process yielding, the sort of "family tree" of Indonesian batik, that is called "The Phylomemetic Batik", borrowing the similar analysis in genetics, "the phylogenetic tree of organisms".

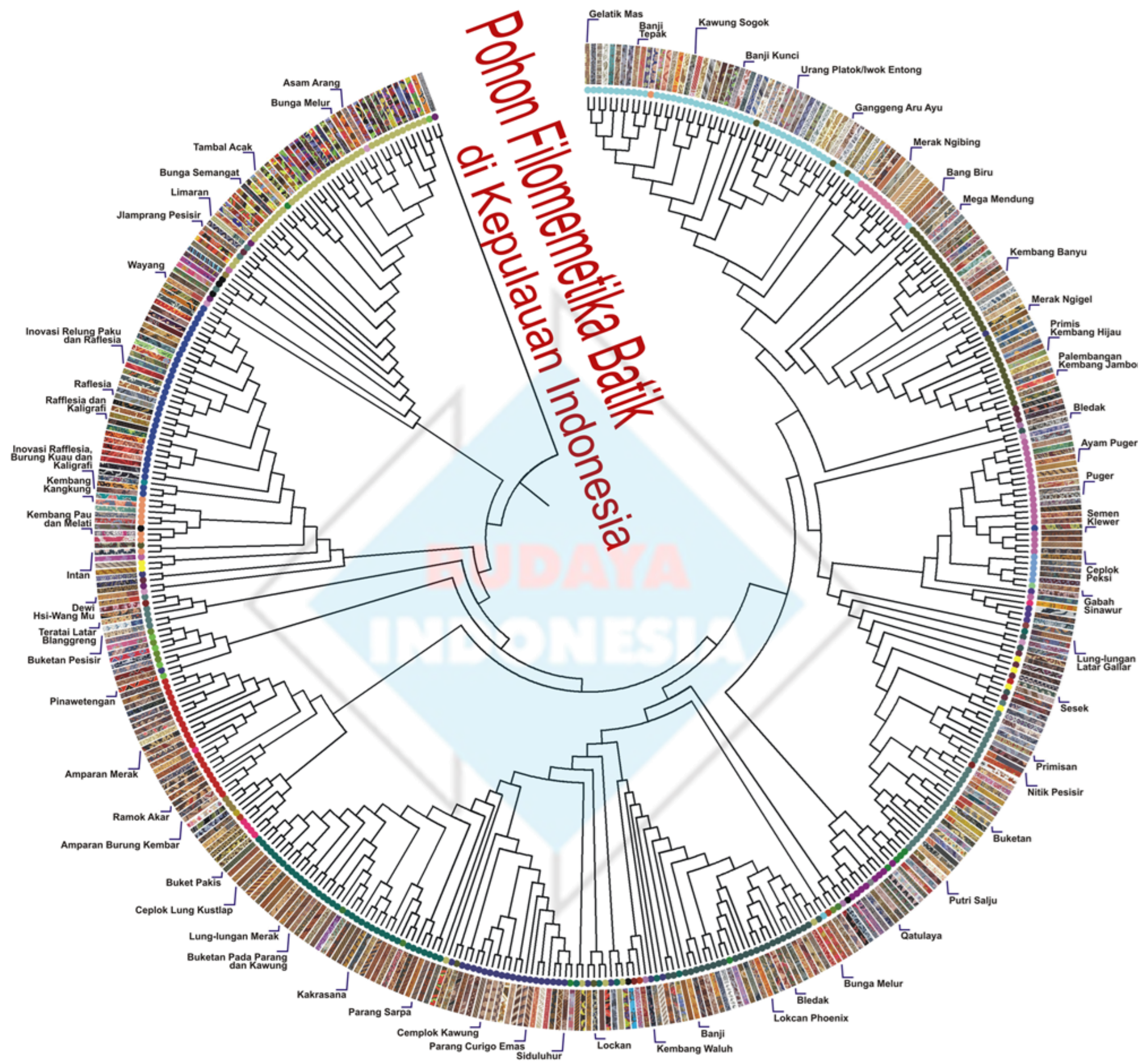

**Figure 5.** The phylomemetic tree of Indonesian batik representing the diversities within Indonesian traditional fabric decorations [19].

Data can also be in the form of texts. Democracy in information era has flooded citizens with so many news channels, including Indonesia. Celebrities, be it political or entertainment, have become public figures with lots of polemical relations among them. The great deal of news channels reported the polemic on daily basis. It is easy to see the polemic between only few actors, but it will not be easy to see when it involves many celebrity actors. When all the statements of all celebrities is digitally archived and treated as corpus, semantic processing come into possibility. Statements conveyed in Indonesian language, $\omega \epsilon \mathbb{L}$, is transormed into sentiment between one actor to another, be it positive sentiment (when two actors agree with one or more issues), negative (conflicting opinion), or neutral,

$$R(\omega \epsilon \mathbb{L}) \equiv \delta_{sentiment}(\omega) \quad (5)$$

Which in detail,

$$\delta_{sentiment}(i,j) = \begin{cases} 1, & if\ w_{v_i v_j} > 0 \\ 0, & if\ w_{v_i v_j} = 0 \\ -1, & if\ w_{v_i v_j} < 0 \end{cases} \quad (6)$$

where $w_{v_i v_j}$ is the weighing factor of sentiment covered by the media,

$$w_{v_i v_j} = \frac{\sum_k v_{ik} . v_{jk}}{\sqrt{\sum_k^N (v_{ik})^2} \sqrt{\sum_k^N (v_{jk})^2}} \quad (7)$$

based upon the sentiment of actor $i$ and actor $j$ to a public topic $k$, writen as $v_{ik}$ and $v_{jk}$ respectively. From this modeling, we have the sentiment-mapping of Indonesian celebrities, summarized from their statements as recorded by journalism [21].

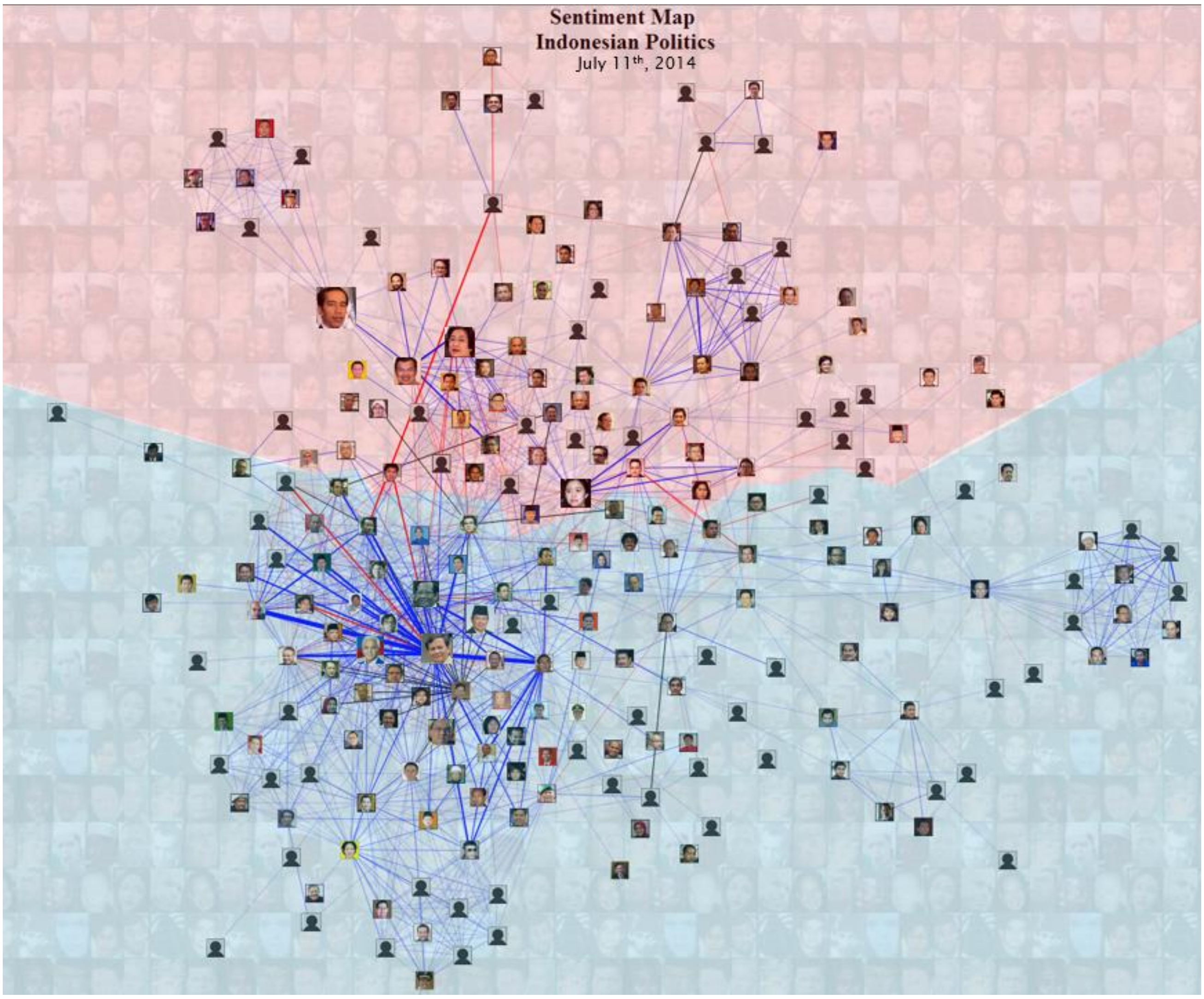

**Figure 6.** The daily Sentiment Mapping among Indonesian political celebrities. The blue line denotes the positive sentiment, the red one negative, and the black one neutral. Due to the political situation heading for General Election 2014, the sentiment among the political figures are halved into two political poles of presidential candidacy.

As demonstrated in figure 6, the Sentiment Mapping from the Newsmedia Processing Suite made it possible to see the clustered celebrities based on some issues being highly reported by the news channels. The abundant information from the news channel has been transformed into new kind of information via a semantic processing.

Nonetheless, our recent information era has made the living generation into the information-generating modern human. Social media has made all people not only consume information (as provided by the press), but also produce information that the media stage does not only belong to celebrities or public figures. There are also sentiments that can be captured by the flowing online status shared publicly by citizens.

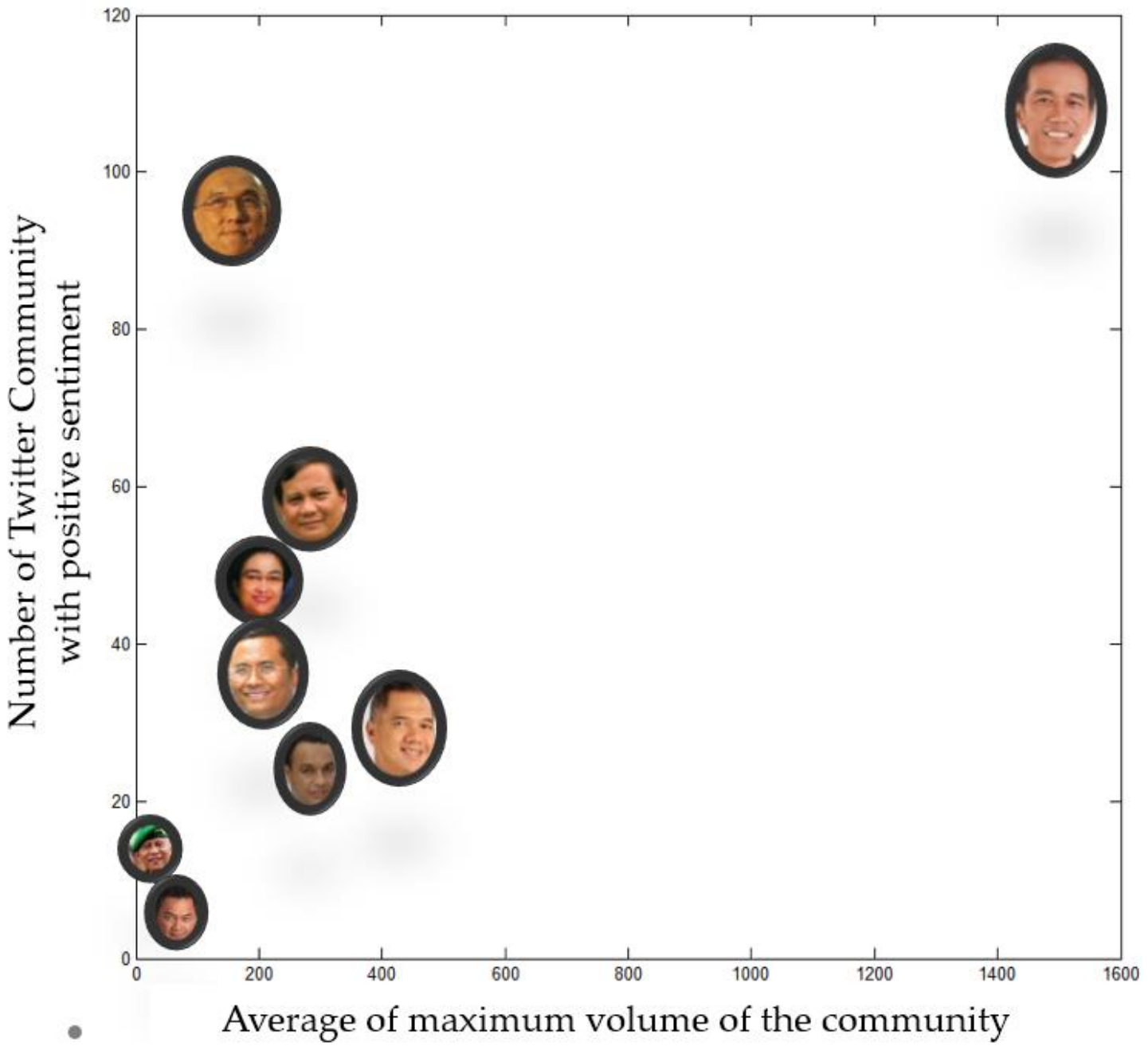

**Figure 7.** The community detection algorithm within tweets about some possible candidates of national figures before Indonesian presidential election in 2014. The more a figure in the content of public conversation with positive sentiment, the larger the chance to win the votes.

In social media, like twitter, people do conversation. They reply, quote, or retweet tweets due to their sentiment about the content of them $(\omega \epsilon \mathbb{L})$. As they do that, actually they are making connection one another [*cf.* 4, 20]. There can be drawn graph of engagement among twitter accounts based on their behavior in the social media. Learning machine detect and represent it in graph via scrutinizing particular twitter account, $(v)$ tweeting, retweeting, and replying behavior, making engagement $(e)$ with other twitter account, of graph as representation,

$$R(\omega \epsilon \mathbb{L}) \equiv G(v, e) \quad (9)$$

Later, analytical computation is delivered for calculating the twitter accounts herding based on similar sentiment over online conversations. Millions of people tweet and they are absorbed in their social network, making the localized communities over topics of conversations. The greater the importance of a topic, the greater number of localized communities over such topics. The volume of the communities varies; depend on the number of twitter accounts play role in each of it,

$$A \circ R(\omega \epsilon \mathbb{L}) \equiv \delta_{sentiment}(tweets(t)) \xrightarrow{A} community_counts \quad (10)$$

Note that the analytical processes, $A(x)$, is delivered as a composition on the data representation, $R(x)$. The community counts on a social topic shows its importance globally. Figure 8 shows some exemplification on the popularity of some national figures considered by citizens to be the next president candidates in the Indonesian General Election 2014. The algorithm of twitter community detection is run long before the Election Day, and the calculation even showed the power of predictability since the result of the election is the political celebrity with the highest community involved in the tweeting engagement with positive responses.

## 4. Closing Remarks

The two operational functions, $A(x)$ and $R(x)$ are elementary functions on how Data Science may deliver new information from the information within the collected data. The interplay between the two functions plays important role in how the emerging Data Science cope with various types of data, be it time-based signals, images, spatial system, and texts. While the global trends of Data Science is related to the trivial terminology of the large data set, named "big data" [9], the existing data due to the vast Indonesian archipelago gives challenge to see the country in the fashion of data.

Reviewing some applications on how visualization or representation of data in Indonesia some interesting views are shown about the country. Cartograms demonstrates how the archipelagic geography may be visualized along with the data on particular statistical aspects. The mapping of Indonesian batik for instance, gives "new kind of" picture of the diversity of Indonesian collective intelligence resembled in hundreds of ethnic groups. The econophysics on Indonesian market data gives insights on the complexity on how people in the country do the economy. The semantic data crunch on Indonesian news media emerges new picture on the interaction of the elite celebrities due to political events nation-wide. Eventually, the social media data crunch gives insight on what and how citizens collectively reflect the nation-wide situations due to the conversational engagement through online status updates. All of them are shown as the interaction (and various composition) between the operational functions to represent and analyze the data which also, are in various forms.

Crunching data is motivated to gain information from the existing information, and some of the depicted demonstrations express some new challenges to know more about the country via data, or even further extend the meaning of the national entity. This is the portrait how Indonesia is embracing the data science.